# Formation of periodic surface structures on dielectrics after irradiation with laser beams of spatially variant polarisation: a comparative study

Antonis Papadopoulos [1,2], Evangelos Skoulas [1,2], George D.Tsibidis [1♣], and Emmanuel Stratakis [1,2♠]

[1] Institute of Electronic Structure and Laser (IESL), Foundation for Research and Technology (FORTH), N. Plastira 100, Vassilika Vouton, 70013, Heraklion, Crete, Greece
[2] Materials Science and Technology Department, University of Crete, 71003 Heraklion, Greece

**ABSTRACT**

A comparative study is performed to explore the periodic structure formation upon intense femtosecond pulsed irradiation of dielectrics with radially and azimuthally polarised beams. Laser conditions have been selected appropriately to produce excited carriers with densities below the optical breakdown threshold in order to highlight the role of phase transitions in surface modification mechanisms. The frequency of the laser induced structures is calculated based on a theoretical model that comprises estimation of electron density excitation, heat transfer, relaxation processes, and hydrodynamics-related mass transport. The influence of the laser wavelength in the periodicity of the structures is also unveiled. The decreased energy absorption for azimuthally polarised beams yields periodic structures with smaller frequencies which are more pronounced as the number of laser pulses applied to the irradiation spot increases. Similar results are obtained for laser pulses of larger photon energy and higher fluences. All induced periodic structures are oriented parallel to the laser beam polarisation.

♣ tsibidis@iesl.forth.gr (Theory)
♠ stratak@iesl.forth.gr (Experiment)

1. Introduction

Laser processing of dielectrics with femtosecond pulses has received remarkable attention due to its important technological applications [1-9]. Regarding the underlying physical processes that lead to surface modification, the predominant emphasis was initially centred on the role of electron density in the induced surface profile through the estimation of the optical breakdown damage threshold (OBT) [10-12]. Nevertheless, a more precise approach requires a direct correlation of the laser beam characteristics, the excited electrons density and the induced thermal effects suggests that morphological changes on the surface of dielectrics should involve a thermal criterion [13-15].

One type of surface modification, the laser induced periodic surface structures (LIPSS) (see [16] and references therein) has been extensively explored assuming linearly polarised laser beams (LPB). Nevertheless, it is important to elaborate on whether other types of polarisation states lead to structures with different morphological profiles. More specifically, laser beams with cylindrical states, namely radial and azimuthal polarization, have gained considerable attention in the past two decades, as the symmetry of the polarization enables new processing strategies [17] with applications in various fields including microscopy, lithography [18], electron acceleration [19], material processing [17, 20, 21] and optical trapping [22]. The so-called cylindrical vector beams (CVB) have been the topic of numerous theoretical and experimental investigations [20, 23-29]. A recent work on nickel irradiation with femtosecond pulses revealed morphological changes for LPB and CVB for various fluences and number of pulses [30]. Given the wide range of applications of laser processed dielectrics, it is of paramount importance to explore the fundamental mechanisms using CVB. To the best of our understanding, no comparative study has been performed to underline the role of the laser beam polarisation state in the features (i.e. periodicity, height, spot size, dependence on laser pulses applied to the laser spot, etc.) of the induced surface morphology.

In this work, we investigate the mechanism of periodic structure formation induced on fused silica upon irradiation with, radially and azimuthally polarised, (RPB and APB, respectively) femtosecond pulses. To evaluate the role of the polarisation state in the LIPSS periodicity, a comparative study is conducted to quantify morphological changes for using CVB laser beams with respect to fluence and number of pulses variation. The role of the laser beam wavelength on the periodicity of the produced structures is also highlighted.

2. Theory

2.1 Electron excitation and energy absorption.

To simulate the excitation mechanism in fused silica, we consider the multiple rate equation model (MRE) [31, 32]

$$\partial_t n_1 = \frac{n_v - n_e}{n_v} PI(E_G) + 2\tilde{\alpha} n_k - W_{1pt} n_1 - \frac{n_1}{\tau_r}$$

$$\partial_t n_j = W_{1pt} n_{j-1} - W_{1pt} n_j - \frac{n_j}{\tau_r}, 1 < j < k, \quad (1)$$

$$\partial_t n_k = W_{1pt} n_{k-1} - \tilde{\alpha} n_k - \frac{n_k}{\tau_r}$$

where $n_e$, $n_v$, are the number densities of the excited electrons, and valence band electrons ($n_v$=2.2×10$^{22}$ cm$^{-3}$), respectively. The last term in the first equation corresponds to free electron decay that is characterised by a time constant $\tau_r$ ($\tau_r$~150fs in fused silica) that leads to a decrease of the electron density. For the sake of simplicity, formation of self-trapped excitons is ignored in Eq.1 [13, 32, 33]. The rest of the terms in Eq.1, *PI*, $\tilde{\alpha}$ and $W_{1pt}$ correspond to the photoionization rate, impact ionization probability and the probability for single photon intraband absorption, respectively [32]. With respect to the intensity of the laser beam *I*, the attenuation of the local laser intensity is determined by multiphoton ionisation and inverse bremsstrahlung absorption assuming a laser beam of frequency $\omega_L$ (i.e. wavelength $\lambda_L$) [13, 14, 34].

$$\frac{\partial I(t,\vec{x})}{\partial z} = -N_{ph} \hbar \omega_L \frac{n_v - n_e}{n_v} PI(E_G) - \alpha(n_e) I(t,\vec{x}) \quad (2)$$

where

$$I(t,x,y,z=0) = (1 - R(t,x,y,z=0)) \frac{2\sqrt{\log(2)}}{\sqrt{\pi}\tau_p} E_d \exp\left(-4\log(2)\left(\frac{t-3\tau_p}{\tau_p}\right)^2\right) F(x,y,z) \quad (3)$$

In Eq.2, $N_{ph}$ corresponds to the minimum number of photons necessary to be absorbed by an electron in the valence band to overcome the relevant energy gap $E_G$ and reach the conduction band. Also, $R(t,x,y,z=0)$ stands for the reflectivity of the material, $E_d$ is the fluence while $F(x,y)$ is proportional to $|\vec{E}|^2$ (as $\vec{E}$ is the applied laser electric field). Hence,

a. For a radially polarized beam, $\vec{E}_r = HG_{10}\hat{x} + HG_{01}\hat{y}$, where $\vec{E}_r$ denotes radial polarization and $\hat{x}$, $\hat{y}$ are the unit vectors along the *x*- and *y*-axis, respectively. Hence, the field is expressed as the superposition of orthogonally Hermite-Gauss $HG_{01}$ and $HG_{10}$ modes with *x*- and *y*-polarisation, respectively, which yields [23, 35]

$$F(x,y) \sim \frac{x^2 + y^2}{(R_0)^2} \exp\left(-\frac{2(x^2+y^2)}{(R_0)^2}\right) \quad (4)$$

where intensity is higher at distance $R_0/\sqrt{2}$ from the spot-centre.

b. On the other hand, for an azimuthal beam, $\vec{E}_\phi = HG_{10}\hat{y} + HG_{01}\hat{x}$, where $\vec{E}_\phi$ denotes azimuthal polarization. Hence, the field is expressed as the superposition of orthogonally Hermite-Gauss $HG_{01}$ and $HG_{10}$ modes with y- and x-polarisation, respectively, which yields $F$ equal to Eq.(4) [23, 35]

The transient optical properties of the irradiated material (i.e. refractive index, extinction and absorption coefficient and reflectivity) are computed through the evaluation of the dielectric constant for materials with a band gap $E_G$ [36]

$$\varepsilon(n_e) = 1 + (\varepsilon_{un} - 1)\left(1 - \frac{n_e}{n_v}\right) - \frac{n_e e^2}{\varepsilon_0 m_r \omega_L^2} \frac{1}{1 + i\frac{1}{\omega_L \tau_c}} \tag{5}$$

where $\varepsilon_{un}$ corresponds to the dielectric constant of the unexcited material, $\varepsilon_0$ stands for the vacuum permittivity, $e$ is the electron charge, $m_r$ (=$0.5m_e$, $m_e$ stands for the electron's mass) is the electron's reduced mass, and $\tau_c$=0.5 fs [37] is the Drude damping time. It is evident that in both types of CVB, $F(x,y)$ is the same, however, the absorbed energy in the material is characterised by the difference of the reflectivity $R$ between RP and AP beams. More specifically, a distinction should be identified on the calculated values of the reflectivity for *s*- or *p*-polarised beams. Hence, since azimuthal (radial) CVB are *s*- (*p*-) polarised, the reflectivity $R$ is larger (smaller) which leads to smaller (larger) energy absorption at increasing angle of incidence (where morphological profile becomes steeper as a result of repetitive irradiation) [38]. To take into account the distribution of the absorbed energy, the Fresnel expressions for the transient reflectivity $R_s$ and $R_p$ for *s*- and *p*-polarised beams, respectively, are directly incorporated in the modelling approach through the following expressions

$$R_s = \left|\frac{\cos(\theta_i) - \sqrt{N^2 - (\sin(\theta_i))^2}}{\cos(\theta_i) + \sqrt{N^2 - (\sin(\theta_i))^2}}\right|^2 \tag{6}$$

$$R_p = \left|\frac{-N^2 \cos(\theta_i) + \sqrt{N^2 - (\sin(\theta_i))^2}}{N^2 \cos(\theta_i) + \sqrt{N^2 - (\sin(\theta_i))^2}}\right|^2 \tag{7}$$

to quantify the effect of the polarisation state. In the above expressions, $N = n + i k_i^r$ is the complex refractive index of the irradiated material ($n$ is the real part of the refractive index and $k_i^r$ is the extinction coefficient) and $\theta_i$ stands for the angle of incidence.

## 2.2 Electron-Lattice relaxation processes.

Due to the metallic character of the excited material, a TTM model is used to describe the spatio-temporal dependence of the temperatures $T_e$ and $T_L$ of the electron and lattice subsystems [10, 32, 39], respectively

$$C_e \frac{\partial T_e}{\partial t} = \vec{\nabla}\left(k_e \vec{\nabla} T_e\right) - g\left(T_e - T_L\right) + S(\vec{x}, t)$$

$$C_L \frac{\partial T_L}{\partial t} = g\left(T_e - T_L\right) \tag{8}$$

The complete expression for source term $S(\vec{x},t)$ is given by [14, 32]

$$S(\vec{x},t) = \left(N_{ph}\hbar\omega - E_G\right)\frac{n_v - n_e}{n_v} PI(E_G) - E_G \tilde{\alpha} \frac{n_k}{n_V} + \alpha(n_e) I(t,\vec{x}) - \frac{3}{2} k_B T_e \frac{\partial n_e}{\partial t} - \frac{3}{2} k_B T_e \frac{n_e}{\tau_r} \tag{9}$$

to account for photoionisation of electrons, avalanche ionisation, free electron absorption and energy loss due to trapping processes. We point out that a term that describes the divergence of the current of the carriers ($\vec{\nabla} \cdot \vec{J}$) has not been taken into account (it appears when carrier dynamics in irradiated silicon is explored upon irradiation [40], however, it turns out that for small pulse durations its contribution is negligible [41]). The temperature, electron density and temporal dependence of the thermophysical properties, $C_e$, $k_e$, $C_L$ and $g$ are provided from well-established expressions derived from free electron gas based on the metallic character of the excited material [13, 32, 42].

## 2.3 Carrier density related LIPSS periodicity.

To correlate the excited electron density with surface structures, the inhomogeneous energy deposition into the irradiated material is computed through the calculation of the product $\eta(k_L,k_i) \times |b(k_L)|$ as described in the Sipe-Drude model [43]. In the above expression, $\eta$ describes the efficacy with which the surface roughness at the wave vector $k_L$ (i.e. normalized wavevector $|k_L|=\lambda_L/\Lambda$, where $\Lambda$ stands for the predicted periodicity due to electrodynamics) induces inhomogeneous radiation absorption, $k_i$ is the component of the wave vector of the incident beam on the material's surface plane and $b$ represents a measure of the amplitude of the surface roughness at $k_L$. To introduce surface roughness, the value $b=0.4$ is initially assumed [32], while the computation of the influence of efficacy influence due to a variable amplitude and shape factor of the produced profile is also considered that accounts for expected corrugation changes upon successive irradiation. To take into account the changes in the periodicity values for AP and RP beams, the computation of $\eta$ is performed for $s$- and $p$- polarized beams, respectively [43, 44]. It is assumed that $k_i$ is along the $\hat{r}$ unit radius vector.

## 2.4 Fluid dynamics

To correlate the induced surface profile with thermal effects, two processes are considered resulting from the intensity profile of the laser beam and the laser energy. In principle, surface modification is usually associated with a phase transition (i.e. melting of the material). Nevertheless, near the centre of the laser spot, the deposition of energy is not large enough to produce a phase transition and, thermoplastic effects are predominantly responsible for the permanent deformations. Due to the, significantly, smaller size of surface deformation in the irradiated spot due to the latter effects [30, 45], the role of elastoplastic effects in the surface modification is ignored in this work.

It is also assumed that the material in the molten phase behaves as an uncompressed fluid and therefore, the Navier-Stokes equation is used to describe the fluid dynamics

$$\vec{\nabla} \bullet \vec{u} = 0$$
$$\left(C_L^{(m)} + \vec{\nabla} \bullet (\vec{u}T_L)\right)\frac{\partial T_L}{\partial t} = \vec{\nabla} \bullet (\kappa_L^{(m)} \vec{\nabla} T_L) \qquad (10)$$
$$\rho_L^{(m)}\left(\frac{\partial \vec{u}}{\partial t} + \vec{u} \bullet \vec{\nabla} \vec{u})\right) = \vec{\nabla} \bullet \left(-P + \mu^{(m)}(\vec{\nabla}\vec{u}) + \mu^{(m)}(\vec{\nabla}\vec{u})^T\right)$$

where $\vec{u}$ is the fluid velocity, $\mu^{(m)}$ is the dynamic viscosity of the liquid, $\rho_L^{(m)}$ is the density of the molten material, $T_{melt}$ is the melting temperature for $SiO_2$, $P$ stands for pressure, $\kappa_L^{(m)}$ and $C_L^{(m)}$ stand for lattice heat conductivity (see Table 1) and lattice heat capacity respectively, of the molten material. The second equation in Eq.10 results from energy conservation requirements. To describe the fluid dynamics, the movement of the isothermal $T_{melt}$ is followed to determine the resolidification process.

## 3. Simulation

To determine the modification characteristics, a solidification criterion is used based on the lattice displacements when the associated lattice temperature drops to lower values than the melting point [30, 32, 46, 47]. To simulate the induced surface modification, on a spot of $R_0$=15μm a three dimensional finite-difference method in a staggered grid is employed [46] to solve numerically the heat transfer equations, phase change and resolidification process [32]. At time $t$=0, both electron and lattice temperatures are set to 300K. For the part of the material which is in the liquid phase non-slipping conditions (i.e. velocity field is zero) are applied on the solid-liquid interface. The grid size taken for simulations is 2nm (vertical dimension) and 2nm (horizontal dimension). The temporal step is adapted so that the stability Neumann condition is satisfied [49]. In the simulations, it is also assumed that a region where the lattice temperature exceeds the values ~$1.8T_b$ (~4000K)[14] (where $T_b$ is the boiling temperature) is ablated [45, 46, 50]. Regarding the adequate description of fluid dynamics, the hydrodynamic equations are solved in the subregion that contains either solid or molten material. To include the "hydrodynamic" effect of the solid domain, material in the solid phase is modelled as an extremely viscous liquid ($\mu_{solid} = 10^5 \mu_{liquid}$), which results in velocity fields that are infinitesimally small. An apparent viscosity is then defined with a smooth switch function to emulate the step of viscosity at the melting temperature. A similar step function is also introduced to allow a smooth transition for the rest of the temperature-dependent quantities (i.e., heat conductivity, heat capacity, density, etc.) between the solid and liquid phases. For time-dependent flows, a common technique to solve the Navier-Stokes equations is the projection method, and the velocity and pressure fields are calculated on a staggered grid using fully implicit formulations. More specifically, the horizontal and vertical velocities are defined in the centers of the horizontal and vertical cells faces, respectively, where the pressure and temperature fields are defined in the cell centers (see description in Tsibidis et al. [46]). Similarly, all temperature-dependent quantities (i.e., viscosity, heat capacity, density, etc.) are defined in the cell centers. While second-order finite difference schemes appear to be accurate for *NP*=1, where the surface profile has not been modified substantially, finer meshes and higher-order methodologies are performed for more complex profiles [51, 52]. Furthermore, techniques that assume moving boundaries (i.e. solid-liquid interface) are employed [53].

Table 1 Simulations parameters for $SiO_2$

| Parameter | Value |
|---|---|
| $C_L$ [$10^6$ Jm$^{-3}$ K$^{-1}$] | 1.6 [42] |
| $C_L^{(m)}$ [$10^6$ Jm$^{-3}$ K$^{-1}$] | Fitting [32, 48] |
| $T_{melt}$ [K] | 1988K [32] |
| σ [N/m] | Fitting [32, 48] |
| $\rho_L^{(m)}$ [Kgr/m$^3$] | Fitting [32, 48] |
| $\mu^{(m)}$ [mPa sec] | Fitting [32, 48] |

## 4. Experimental protocol

Commercial polished samples of $SiO_2$ of 99.9% purity and average thickness of 1 mm were used. An Yb:KGW laser source was used to produce linearly polarized pulses with pulse duration equal to 170 fs, 1 kHz repetition rate, at 1026 nm central wavelength and 56μm Gaussian spot diameter. Beams with radial and azimuthal polarisation were generated using an *s*-waveplate. Following the *s*-waveplate, the beams were focused on the sample via an achromatic convex lens of 100mm focal length. The samples were fixed onto a 3-axis motorized stage and positioned perpendicular to the incident beam. The number of pulses was controlled by an electromagnetic beam shutter and all irradiations were performed in ambient environment. The morphology of the laser-induced structures has been characterized by scanning electron microscopy (JEOL JSM-7500F). The morphological features of the structures were computed by a two dimensional fast Fourier transform analysis of the respective SEM images using the Gwyddion software (See Supplementary Material for a more detailed description). The experimental investigation for the period analysis of the irradiated laser spots was conducted for multiple experiments, for both azimuthal and radial polarization states. In particular, the presented measured period values are not a product of a single experiment, but of five different experiments, performed under identical conditions. Moreover, for each experiment , at least four measurements of the periodicity at different areas of each spot have been obtained. The standard deviation and the mean of the error bars, presented in the discussion, are the outcome of those measurements.

## 5. Results and discussion

As repetitive irradiation of a material leads to a gradually deepening of the affected zone (an initially flat profile is transformed in a curved morphology that suggests that an increase of the irradiation pulses produce a larger incidence angle), the need to distinguish the influence in the energy absorption of the differently polarised beams, APB and RPB, suggests that it is imperative to estimate the reflectivity of the material at various angles of incidence (up to the Brewster's angle that is approximately $55^0$ for fused silica) as a spatially variable energy absorption will influence the overall density of the excited carriers. Hence, a generic scheme is developed to determine the amount of energy that is absorbed from each segment of the curved geometry as a result of exposure of the material to a sequence of laser pulses [45, 46]. The solution of Eqs.1-7 yields the spatio-temporal distribution of the produced excited carriers. The induced maximum carrier density and the associated reflectivity values for RPB and APB as a function of the angle of incidence are presented in Fig.1 (see also Supplementary Material). The simulation parameters are $E_d=7$J/cm$^2$, $\tau_p=170$fs, $\lambda_L=513$nm and 1026nm. It turns out that the produced excited carrier densities for RPB increases at larger values of $\theta_i$ (Fig.1a) that leads to a decreased reflectivity and therefore enhanced energy absorption from the irradiated material (Fig.1b). It also appears that for the above simulation parameters (i.e. that leads to excitation of relevantly small carrier densities [37]), the influence of the laser wavelength does not yield significant changes in the reflectivity (Fig.1b) despite the expected increase of the excited carrier density at smaller $\lambda_L$ resulting from the larger photon energy. Nevertheless, a focus on an enlarged area (see Supplementary Material) illustrates that for both $s$- and $p$-polarised beams, the induced reflectivity becomes higher at smaller laser beam wavelengths.

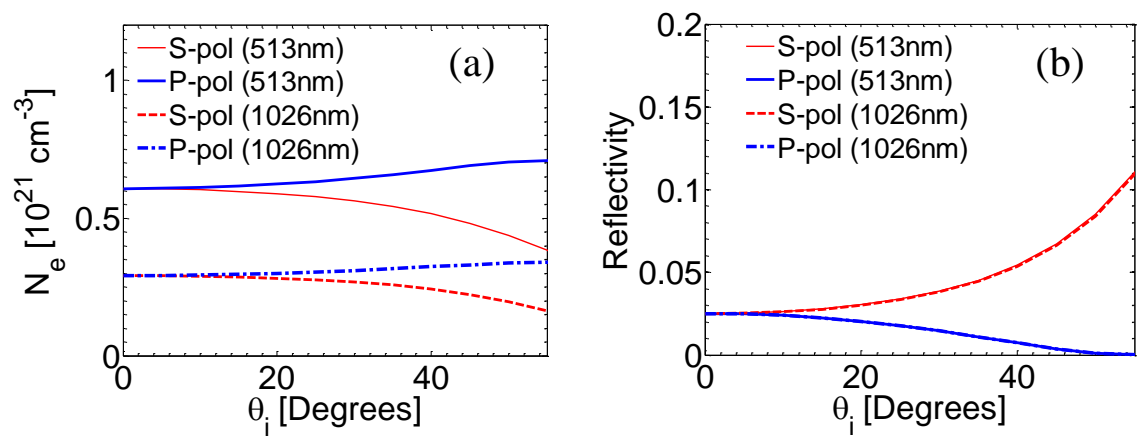

FIG. 1. (Color online) (a) Carrier density and (b) Reflectivity at $z=0$, $x=y=R_0/\sqrt{2}$, for $s$- and $p$- polarisation as a function of $\theta_i$. ($NP=1$, $\tau_p=170$fs, $E_d=7$J/cm$^2$).

On the other hand, the produced excited electron densities are always lower than the OBT $n_{ob}$ (i.e. $n_{ob}=1.06\times10^{21}$cm$^{-3}$ and $4.25\times10^{21}$cm$^{-3}$ at $\lambda_L=1026$ nm and $\lambda_L=513$nm, respectively) which in previous works have been related to the damage threshold [10-12]. Simulation results for $\theta_i=0$ (similar results follow for other angles) illustrate the evolution of the maximum electron density inside the bulk (z-axis at $x=y=R_0/\sqrt{2}$ where the absorbed energy is maximum) (Fig.2a,b). Nevertheless, herein, a thermal criterion is used to introduce surface damage which is demonstrated through the development of lattice temperatures larger than the melting temperature of fused silica (~1988K). As seen in Fig.2c,d, the spatio-temporal lattice temperature shows that a substantially large part of the material has undergone a phase transition. More specifically, there are two regions that require special attention, one that is related to ablation ($T_L>4000$K) and a second with $T_{melt}<T_L<4000$K that describes material in a molten phase. While at $\lambda_L=513$nm, the two phenomena occur in the regions $z\leq290$nm and $290\text{nm}\leq z\leq1.268\mu$m (Fig.2c), respectively, for $\lambda_L=1026$nm a less pronounced damaged region is produced at $z \leq 150$nm and 150nm$\leq z \leq 1.228\mu$m (Fig.2d) as a result of a smaller photon energy.

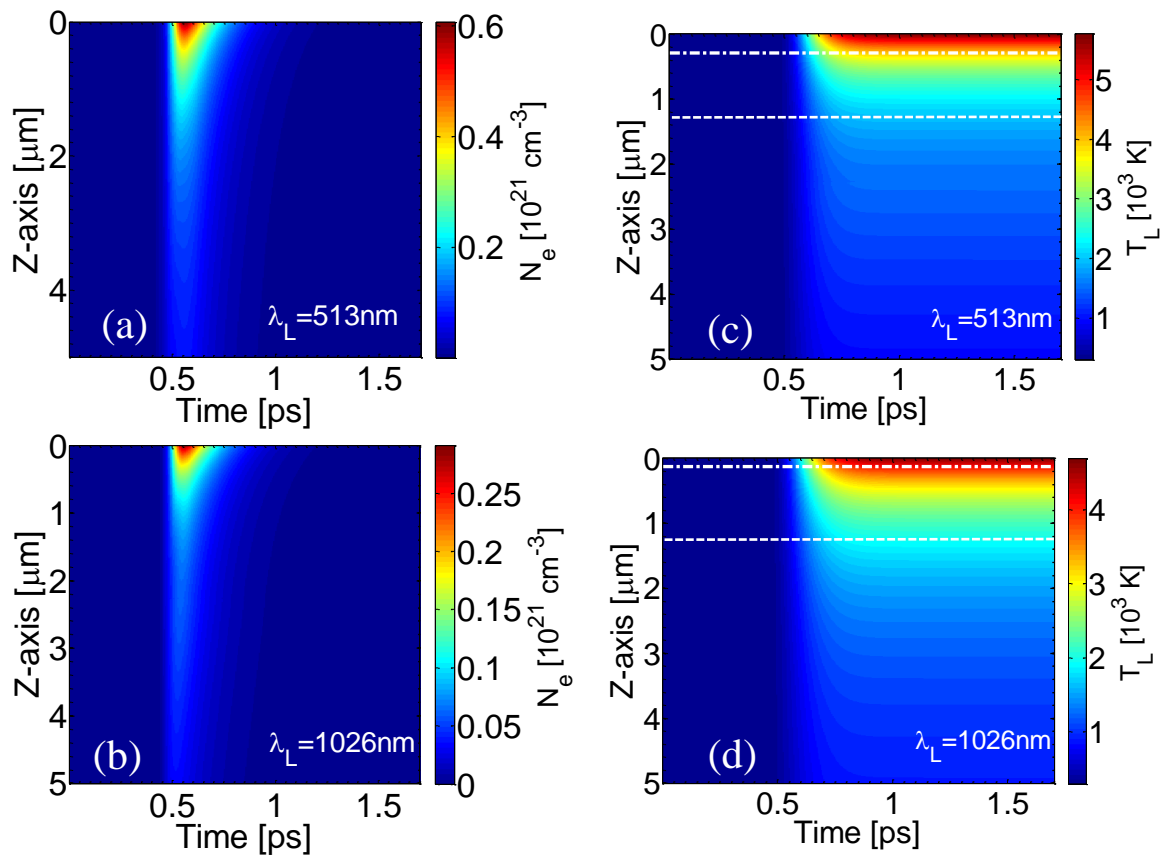

FIG. 2. (Color online) Evolution of maximum free electron density at $x=y=R_0/\sqrt{2}$ for (a) $\lambda_L=513$nm and (b) $\lambda_L=1026$nm (distance $R_0/\sqrt{2}$ from the spot centre). Evolution of maximum lattice temperature at $x=y=R_0/\sqrt{2}$ for (a) $\lambda_L=513$nm and (b) $\lambda_L=1026$nm (distance $R_0$ from the spot centre) ($NP=1$, $\tau_p=170$fs $E_d=7$J/cm$^2$). In (c) and (d) *dashed-dotted* and *dashed* lines indicate ablation and melting depths, respectively.

The differences in the thermal effects for the two polarisation states and the significance of the angle of incidence (up to the Brewster's angle) are also depicted in Fig.3. The theoretical predictions indicate that the induced maximum lattice temperature for RPB is higher than the value computed for APB which is also in agreement with the trend the excited carrier densities follow (Fig.1a). This observation also underlines the more dominant role of *p*-polarisation in the production of thermal effects. On the other hand, the decrease of $T_L$ difference between the values for irradiation with 513nm and 1026nm laser beam for APB (compared to RPB) with increasing angle of incidence can be attributed to the increase of the reflectivity if *s*-polarised waves are used.

Simulations show, also, that due to the induced larger lattice temperatures a deeper damaged region is produced for irradiation with RPB compared to APB. In Fig.4, it is shown that irradiation with a laser beam of $\lambda_L=1026$ nm at an angle $\theta_i=30^0$, produces a molten volume of depth equal to 963 nm (for RPB) while for APB it is smaller (759 nm). Similar conclusions can be deduced at other angles and $\lambda_L$.

Theoretical models and experimental observations have indicated that upon repetitive irradiation periodic structures are formed [30, 32, 45-47, 54, 55]. To present a mechanism that explains the development of periodic structures on fused silica for RPB and APB, simulations are carried out firstly to correlate the spatio-temporal distribution of the excited electrons with the anticipated LIPSS periodicity based on Sipe's theory [43]. The computations have been performed to estimate, also, the role of the angle of incidence and how the previously calculated carrier density and dielectric constant influences the average periodicity of the induced structures. The orientation and size of rippled structures are predicted by computing the efficacy factor for APB (Fig.5) and RPB (Fig.6) for $\theta_i=0^0$, $10^0$, $30^0$, respectively, (similar conclusions can be drawn for other values of $\theta_i$). The efficacy factor cross-line along $k_r$ (for *s*-polarised APB) and $k_\theta$ (for *p*-polarised RPB) provide an estimate of the ripple periodicity due to electrodynamics. Periodicity changes are due to the variation of the electron density resulting from differences in reflectivity, energy absorption and curvature of the irradiated profile. Results for RPB and APB indicate low spatial frequency periodic structures (LSFPS)

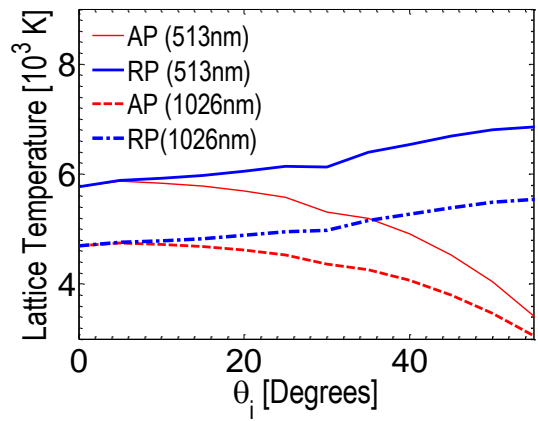

FIG. 3. (Color online): Maximum Lattice temperature dependence on polarisation and angle of incidence.

are formed with an orientation always *parallel* to the polarisation of the electric field on the plane of incidence (Fig.5,6). A similar orientation was also predicted for linearly polarised beams [32, 37]. Results shown in Figs.5,6 are similar for $\lambda_L$=513nm and 1026nm as $k_r$, $k_\theta$ (the components of the LIPSS wave vectors $k_L$ along the *r*- and *θ*- axes) are normalised with the wavelength (the computed different carrier density does not produce substantially different shapes of the efficacy factor distribution). Interestingly, the efficacy factor field in the *k*-space resulting from the *s*-polarised beam does not exhibit a similar symmetry as in the case of the RPB that emphasises on the significant influence of the incidence angle. A similar behaviour has been observed in

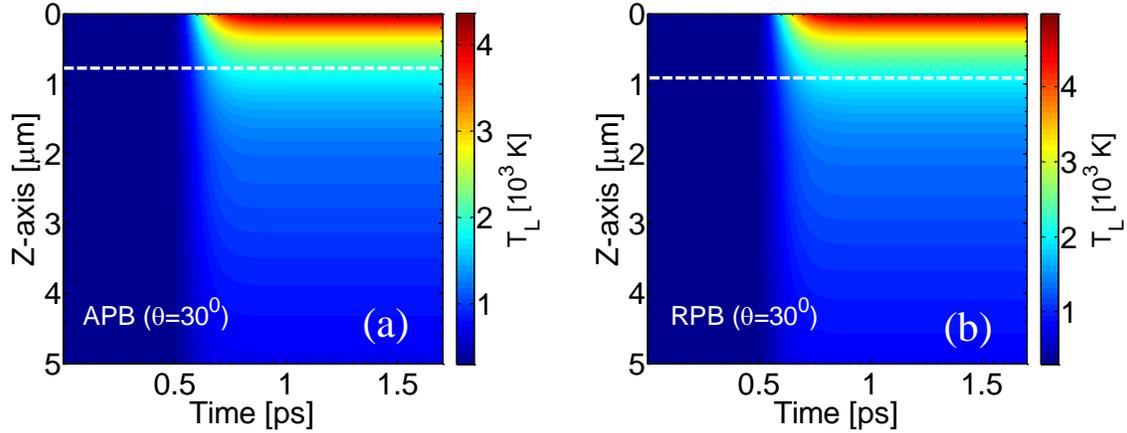

FIG. 4. (Color online): Evolution of maximum lattice temperature at $x=y=R_0/\sqrt{2}$ for (a) APB and (b) RPB (distance $R_0$ from the spot centre) ($NP$=1, $\theta_i$=30$^0$, $\tau_p$=170fs, $\lambda_L$=1026nm, $E_d$=7J/cm$^2$). *Dashed* lines indicate maximum melting depths.

previous studies in semiconductors [44]. The dependence of the computed periodicities derived from the relation of the efficacy factor and the incidence angle is illustrated in Fig.7. The presence of sharp points in the efficacy factor curves (Fig.5b,6b) indicate a remarkably strong absorption which therefore leads to the formation of LIPSS. It is shown that for both $\lambda_L$=513nm and 1026 nm, there are two types of periodic

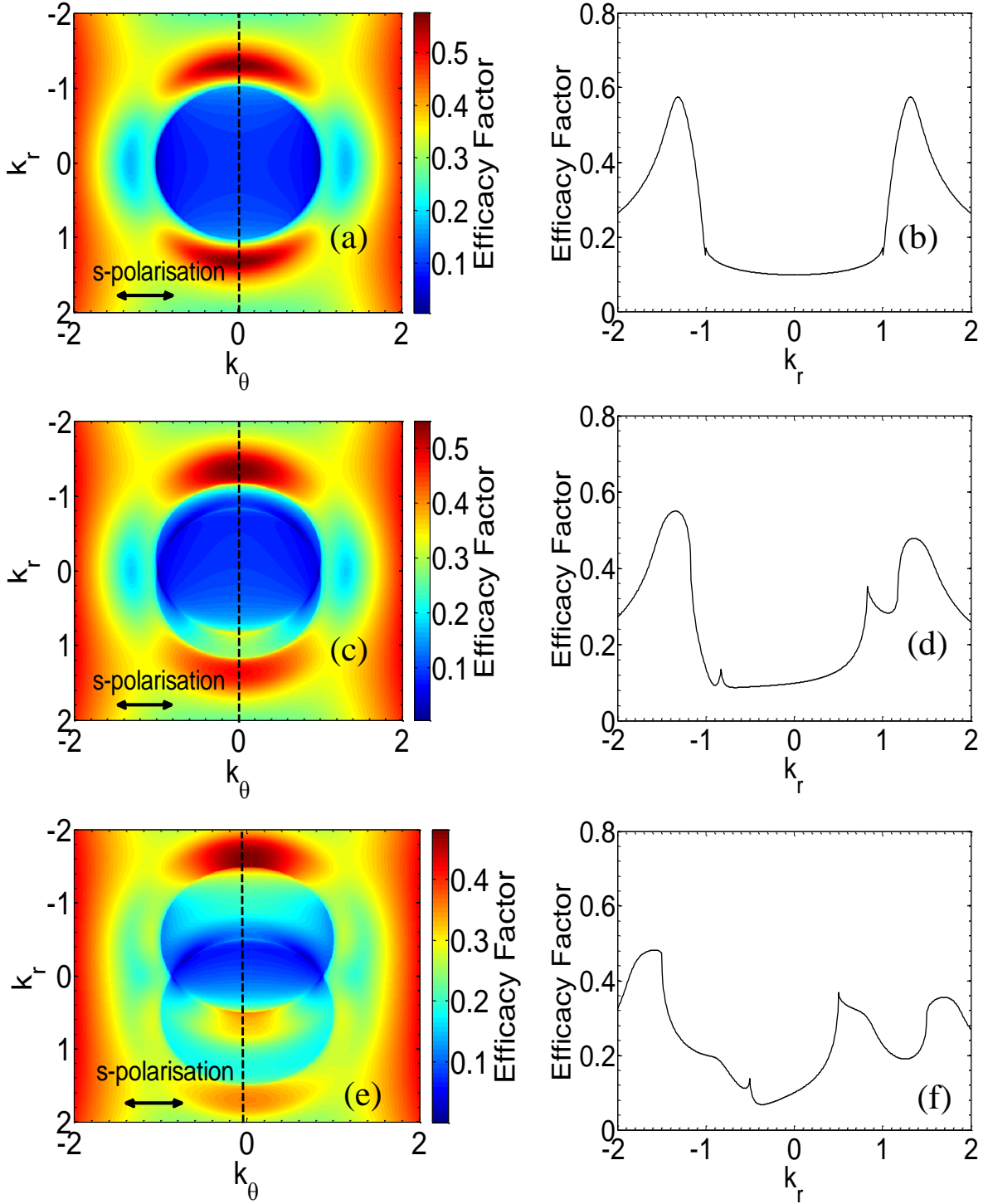

FIG.5. (Color online): APB: (a) Efficacy factor computation as a function of the components of $k_L$ (for $n_e=0.2913\times10^{21}$ cm$^{-3}$, $\theta_i=0^0$). (b) Efficacy factor along the black dashed line in (a). (c) Efficacy factor computation as a function of the components of $k_L$ (for $n_e=0.29128\times10^{21}$ cm$^{-3}$, $\theta_i=10^0$). (d) Efficacy factor along the black dashed line in (c). (e) Efficacy factor computation as a function of the components of $k_L$ (for $n_e=0.3132\times10^{21}$ cm$^{-3}$, $\theta_i=30^0$). (f) Efficacy factor along the black dashed line in (e). ($\lambda_L=1026$ nm).

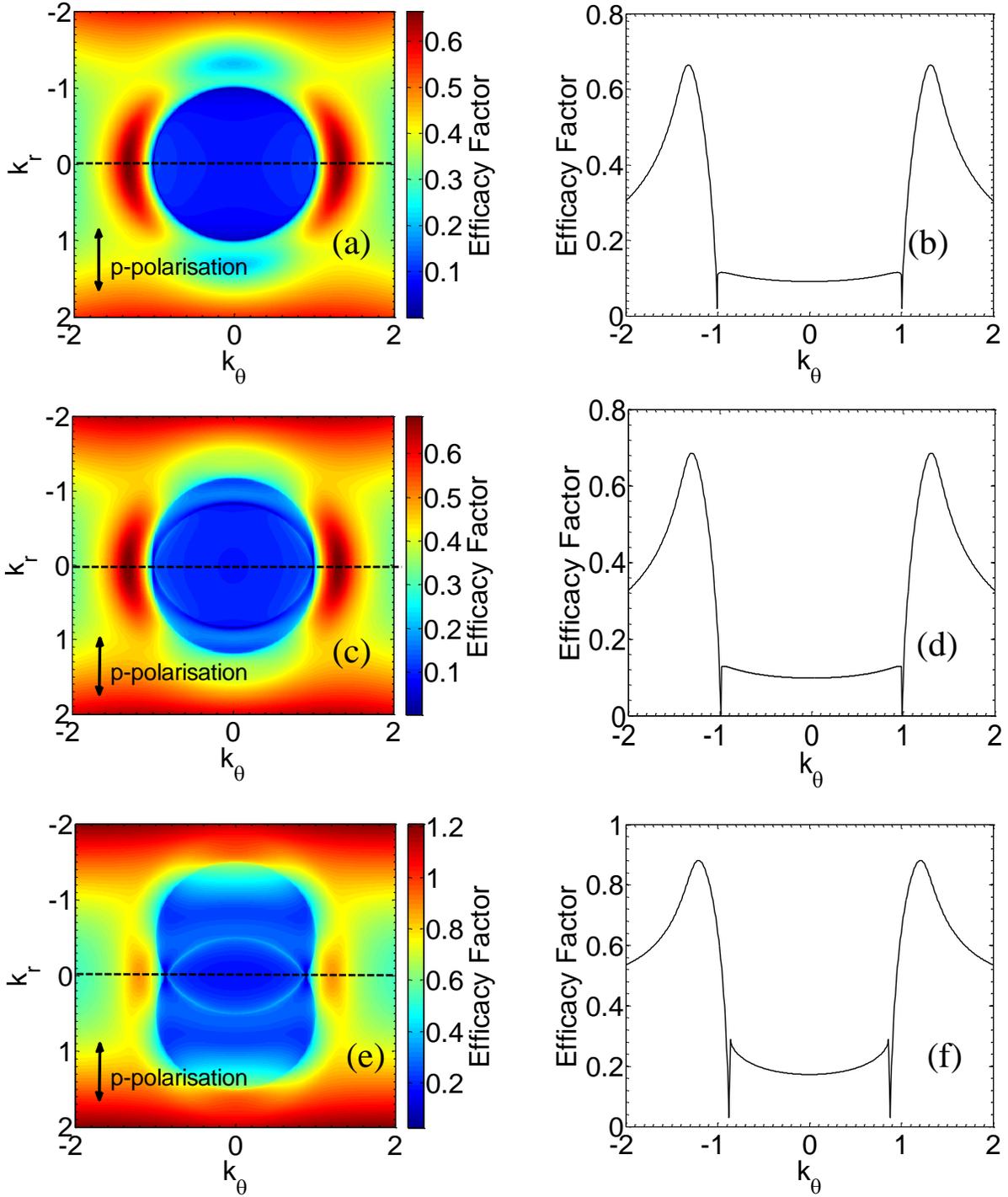

FIG.6. (Color online): RPB: (a) Efficacy factor computation as a function of the components of $\boldsymbol{k_L}$ (for $n_e=0.2913\times10^{21}$ cm$^{-3}$, $\theta_i=0^0$). (b) Efficacy factor along the black

dashed line in (b). (c) Efficacy factor computation as a function of the components of $k_L$ (for $n_e=0.2933\times10^{21}$ cm$^{-3}$, $\theta_i=10^0$). (d) Efficacy factor along the black dashed line in (c). (e) Efficacy factor computation as a function of the components of $k_L$ (for $n_e=0.2932\times10^{21}$ cm$^{-3}$, $\theta_i=30^0$). (f) Efficacy factor along the black dashed line in (e). ($\lambda_L$=1026 nm).

structures that are formed; more specifically, subwavelength (Fig.7a) or suprwavelength (Fig.7b) sized structures are formed at larger angles. The experimental results shown in the discussion (Fig.8 and periodicity measurements in Fig.9) suggest that the second type of periodic structures is preferential.

To correlate the periodic structure formation with the development of thermal effects, the ablation (by removing the lattice points with $T_L$>4000K), carrier density, relaxation processes, phase transformation, Marangoni effects and resolidification process are taken into account for every laser pulse. This is due to the fact that hydrodynamics is expected to contribute to the change of periodicity size (correction to the value $\Lambda$ derived from the efficacy factor computation) of the final profile due to fluid movement [45, 56, 57]. In previous studies, where LSFPS were assumed to be formed through a surface plasmon (SP) excitation mechanism, a periodic function in the source term $S$ in Eq.8 was incorporated that corresponded to the grating period of the structure [30, 46]. As a result, a spatially periodic carrier density, $T_e$ and $T_L$ were produced that led to a similarly periodic thermal response of the lattice system. By contrast, the simulation parameters used in this work yield carrier densities substantially lower than the value that would allow SP excitation in dielectrics (~$10^{22}$ cm$^{-3}$) which also conforms with experimental studies as such modes have never been observed in dielectrics [37]. Nevertheless, to evaluate the frequency of the produced structures upon repetitive irradiation, a spatially periodic function was introduced based on the value computed by the efficacy factor calculation. Hence, the carrier densities, the inhomogeneous energy deposition into the irradiated material, the produced new efficacy factor-based periodicity ($\Lambda$) and the contribution of the thermal effects were computed to estimate the periodicity at the new *NP* after the material has resolidified.

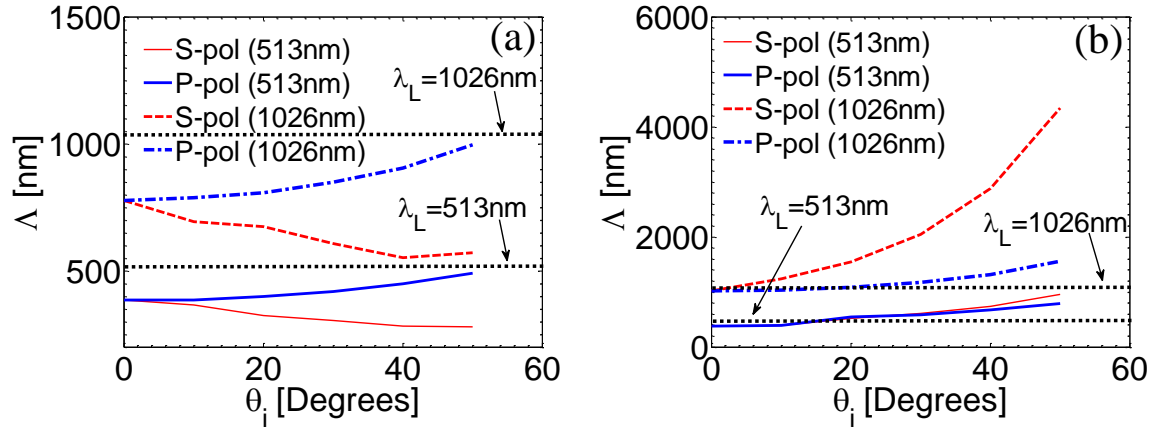

FIG. 7. (Color online): Computed periodicities $\Lambda$ as a function of the incidence angle for structures with (a) large and (b) small $\Lambda$, respectively. ($E_d$=7J/cm$^2$).

The aforementioned scheme towards solving Eqs.1-10, allows a parametric investigation of the morphological features of the produced structures as a function of the laser wavelength, polarisation state, fluence, and number of pulses. In Fig.8a,b, simulation results are illustrated that show the induced morphology after irradiation with APB and RPB, respectively, for *NP*=5. On the other hand, SEM images for the two polarisation states are shown (Fig.8c,d) to provide a qualitative comparison. Notably, similar SEM images were produced using *tightly focused* beams (in the present work, the beams were not tightly focused) in a previous work in which the simulated intensity distribution was produced assuming longitudinal and transverse electric field for the beams [29]. In that work, the estimation of the magnitude of the electric fields was based on their values for RPB (i.e. strong longitudinal component) and APB, respectively [23]. In the present work (where the role of phase transition was particularly highlighted towards determining the final morphological profile), simulations show that periodic structures are formed either perpendicularly to the *r*-axis (APB) or the *θ*-axis (RPB) and they are more pronounced inside the crater (around $R_0/\sqrt{2}$) where the energy deposition is higher. It is noted that, firstly, mass displacement [46], and, secondly, stress-related effects (ignored in these simulations) that are generated around the spot centre are expected to produce an elevated peak the *x=y=0* region that leads eventually to the formation of a 'Mexican-hat' [30] (i.e. the colorbar shows there is a structure formed above the initially flat surface). On the other hand, the simulated rippled morphology for RPB is presented in Fig. 8b, which corresponds to the white rectangular regions A and B in Fig.8d (where the energy deposition is higher according to the spatial distribution of fluence). It is noted that simulation results do not produce a similar (stable and periodic) profile as the respective SEM image for RPB at small radii from the centre of the spot (Fig.8d). This is attributed to strong hydrodynamical forces that do not allow the formation of a regular profile for radii smaller than $R_0/\sqrt{2}$. Furthermore, it is possible that the induced enhanced turbulence that has been observed in those regions require a particular simulation approach to predict a stable and symmetrical profile at smaller radii, however, this is beyond the scope of this work. Nevertheless inside the two (perpendicularly to each other) boxes A and B the ripples are locally parallel to each other. Notably, in Fig.8b the orientation (locally) of the laser beam polarization vector is denoted by the *black doubled* vectors (the notation should not be confused with the direction of a linearly polarised beam). As illustrated in Fig.8b, the height profile distribution in A and B are symmetrical (a more detailed analysis of the height profile is illustrated in the Supplementary Material). To estimate the periodicity, two subsequent peaks of the rippled structures (maxima across the *Z*-axis in the *black* boxes in Fig.8b which correspond to the hills) were considered and the distance between them represent the periodicity of the structure. It is noted that the periodicities were computed for two subsequent structures around the position of maximum energy deposition. The computed periodicities for the two distinct polarisation states are 1023nm (for APB), and 990nm (for RPB), on average, respectively. It is noted that the maximum depth around the region where the energy deposition is higher, is larger for RPB that is justified by the induced larger lattice temperature.

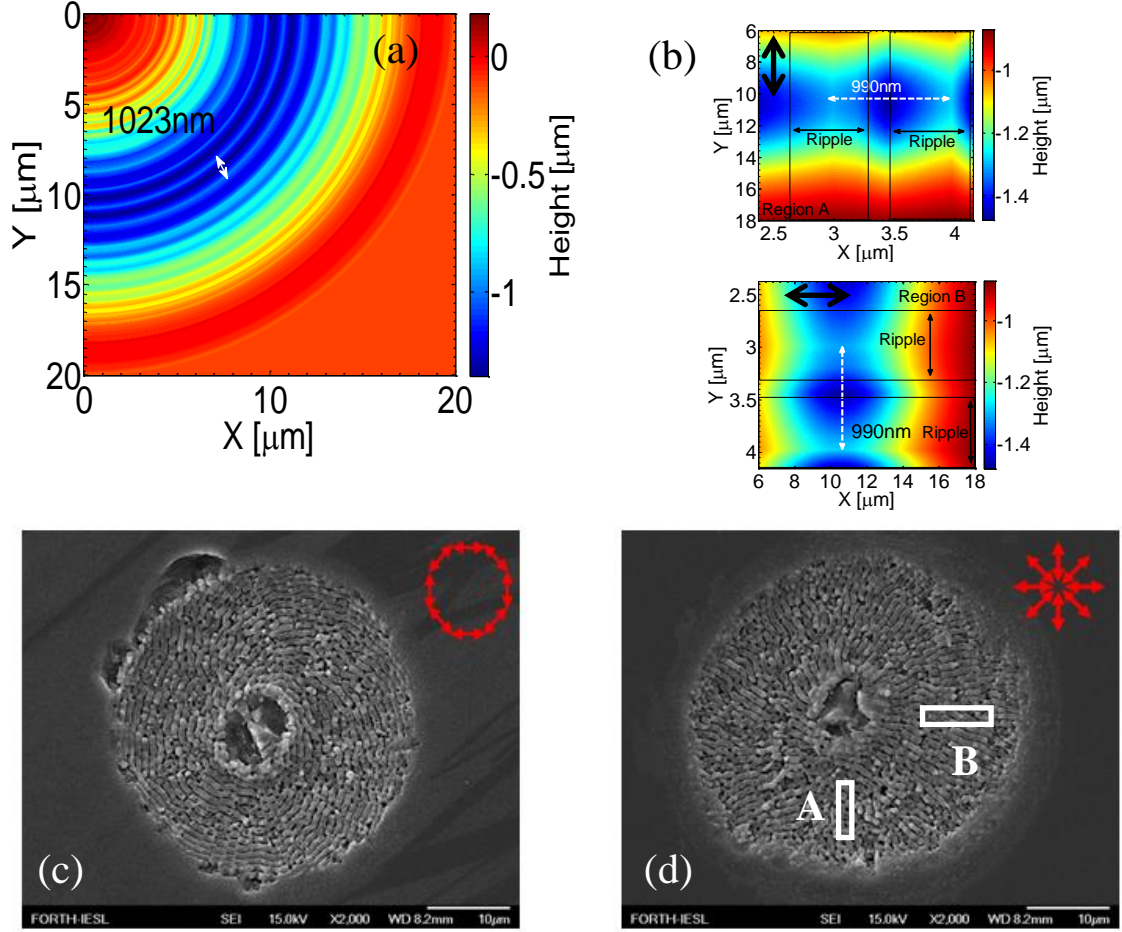

FIG. 8. (Color online): (a) Surface profile for APB (one quadrant). (b) Surface profile for RPB (in *white* rectangular boxes A and B in (d)). SEM image for *NP*=5 for: (c) APB and (d) RPB. ($E_d$=7J/cm$^2$, $\lambda_L$=1026nm). In (b), *Black* doubled vectors indicate the direction of the polarisation vector (locally) while the *white* doubled arrows in the rippled region indicate the periodicity.

Results for the periodicity change as a function of the number of pulses for $\lambda_L$=513nm and $\lambda_L$=1026nm for $E_d$=7J/cm$^2$ show that the ripple periodicity increases with increasing *NP* for APB in contrast to the periodicity due to RPB (Fig.9a). Although repetitive irradiation leads to a deeper profile and a larger angle of incidence for RPB compared to APB the produced carrier densities influence the efficiency of the energy absorption of the energy and leads to a shift of the sharp points of the efficacy factor to smaller $k_L$ for APB and therefore, larger wavelengths (Fig.7b) [37]. To evaluate the influence of the energy deposited on the material, a similar approach was ensued by keeping the number of *NP* constant (*NP*=20) while the fluence remained constant. It is noted that the periodicity of the produced periodic structures as a function of the fluence follows a similar monotonicity with the one described in the previous paragraph. This behaviour is also attributed to the carrier density increase at larger $E_d$ that produced a more efficient energy absorption at smaller $k_L$ (and therefore larger periodicities).

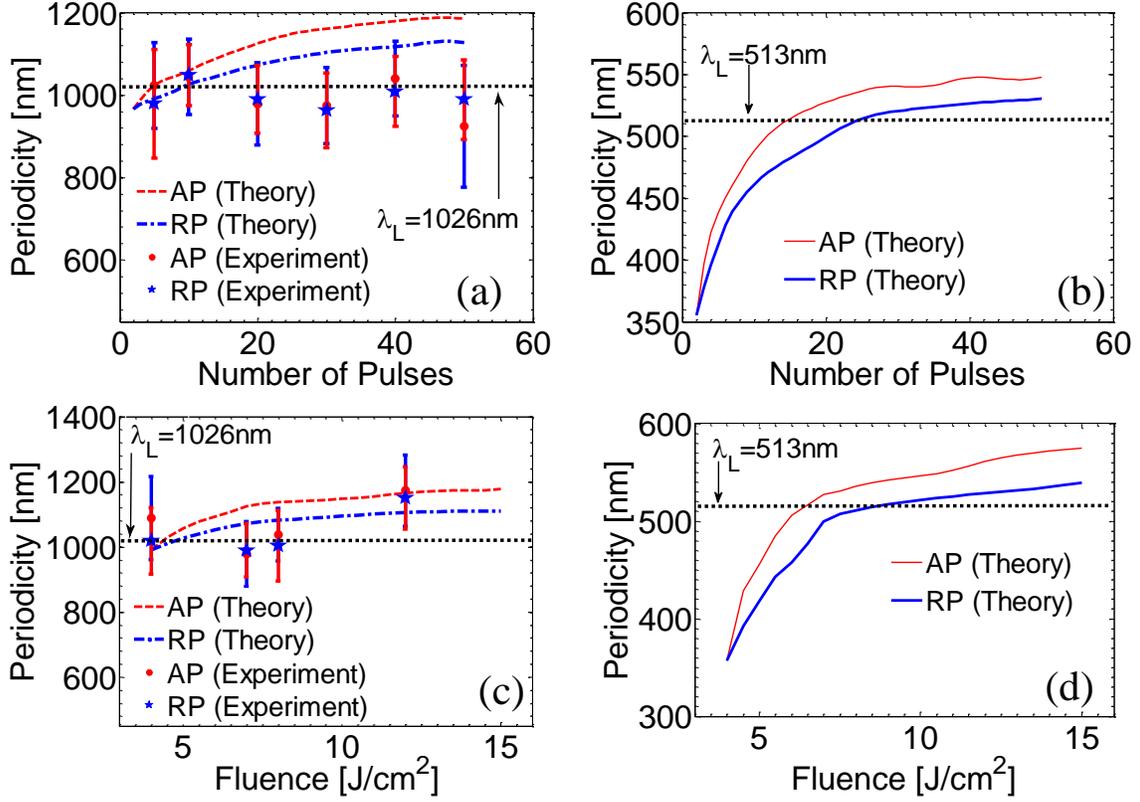

FIG. 9. (Color online): Periodicity of the laser-induced structures as a function of $NP$ (a) for $\lambda_L$=1026nm, (b) for $\lambda_L$=513nm for $E_d$=7/cm$^2$. Periodicity of the laser-induced structures as a function of fluence (c) for $\lambda_L$=1026nm, (d) for $\lambda_L$=513nm for $NP$=20.

To validate quantitatively the theoretical results, data based on experimental observations are also illustrated in Fig.9 for irradiation with laser beam of $\lambda_L$=1026nm (SEM images for $\lambda_L$=513nm were difficult to analyse due to the small size periodicity and therefore experimental data for this wavelength have not been included). It is evident that although the theoretical model predicts a small variation of the periodicities for APB and RPB, the experimental estimation of periodicity includes a large error which does not allow a conclusive physical interpretation on which polarisation type produces larger periodicities and structures. By contrast, the proposed theoretical mechanism correlates a distinct hierarchy of the periodicity values with respect to the polarization state (Fig.9a,b). This tendency is very interesting as it can reveal special features of processing of materials by using laser beams of various polarization types. Although, the discrepancy of the periodicities for the two cases is difficult to be distinguished experimentally, the trend can be sufficiently explained by the differences in the efficiency of the energy absorption and therefore the density of the excited carriers; hence, periodicity size for APB irradiation should lean above that which is produced using RPB (Fig.9a,b). It is evident that at decreasing angle of incidence, the difference between the produced periodicities due to the two polarisation states disappears as there is no energy absorption difference. Finally, the periodicity values appear to saturate at larger values of fluence and *NP* (Fig.9c,d) which has also been observed theoretically and experimentally in previous studies [37, 46].

To summarise, the proposed multi-scale methodology provides new insights into the mechanism that characterises laser-matter interaction as it can allow a parametric investigation of influence the laser beam parameters on the features of the induced morphological changes on the irradiated material. The approach emphasises on the role of the laser beam polarisation in the modulation of the morphological features of the irradiated material and it can provide a recipe for a systematic analysis of the arising possibilities in ultrafast laser-based micro- and nano-fabrication. It should be noted that in addition to the differences between the periodicities values (at these laser energies) of APB and RPB, one aspect that is also significant to underline is that the two polarisation states produce periodic structures with different orientation. Hence, a combination of these (or even more complex) states could be used to produce interesting biomimetic structures [27, 28] that can be used in a wide range of applications.

## 6. Conclusions

In conclusion, we have performed a comparative study to explore and interpret the surface profile and the periodicity of the self-assembled periodic structures formed upon irradiation of fused silica with CVB femtosecond laser pulses at two laser beam wavelengths. It was shown that compared to RPB, APB lead to periodic structures with a larger periodicity than that produced with APB and this difference is more enhanced as the irradiation pulses/fluence increase. On the other hand, the two polarisation states lead to the formation of periodic structures with different orientation. This significant conclusion emphasises particularly the ability to control the size/orientation of the morphological changes via modulating the beam polarization; it is evident that, laser processing through control of use of beams with various polarisation states may provide novel types of surface and bulk structures with significant advantages for potential applications.

## Acknowledgement

This work has been supported by the project *LiNaBioFluid*, funded by EU's H2020 framework programme for research and innovation under Grant Agreement No. 665337 and from *Nanoscience Foundries and Fine Analysis* (NFFA)–Europe H2020-INFRAIA-2014-2015 (Grant agreement No 654360). Funding is also acknowledged from the General Secretariat for Research and Technology (GSRT) and Hellenic Foundation for Research and Innovation (HFRI), No. 130229/I2.## References